\input{aipcheck}

\documentclass[
    ,final            
  ]
  {aipproc}

\layoutstyle{8x11double}
\usepackage{graphicx}
\usepackage{amsbsy}

\begin{document}

\title{New Relativistic Particle-In-Cell Simulation Studies of Prompt and Early Afterglows from GRBs}

\classification{Relativistic Particle-In-Cell Simulation, Studies of Prompt and Early Afterglows from GRBs}
\keywords      {Weibel instability, magnetic field
generation, radiation}

\author{K.-I. Nishikawa}{
  address={National Space Science and Technology Center,
  Huntsville, AL 35805}
}

\author{J. Niemiec}{
address={Institute of Nuclear Physics PAN, ul. Radzikowskiego 152, 31-342 Krak\'{o}w, Poland}}

\author{H. Sol}{
  address={LUTH, Observatore de Paris-Meudon, 5 place Jules Jansen, 92195 Meudon Cedex, France} }

\author{M. Medvedev}{
  address={Department of Physics and Astronomy, University of Kansas, KS
66045, USA} 
}

\author{B. Zhang}{
  address={Department of Physics, University of Nevada, Las
Vegas, NV 89154, USA} 
}

\author{\AA. Nordlund}{
address={Niels Bohr Institute, University of Copenhagen, 
Juliane Maries Vej 30, 2100 Copenhagen \O, Denmark}
}

\author{J. Frederiksen}{
 address={Niels Bohr Institute, University of Copenhagen, 
Juliane Maries Vej 30, 2100 Copenhagen \O, Denmark}
}

\author{P. Hardee}{
  address={Department of Physics and Astronomy,
  The University of Alabama,
  Tuscaloosa, AL 35487} }

\author{Y. Mizuno}{
  address={National Space Science and Technology Center,
  Huntsville, AL 35805}}

\author{D. H. Hartmann}{
  address={Clemson University, Clemson, SC 29634-0978, USA}
}

\author{G. J. Fishman}{
  address={NASA/MSFC,
  Huntsville, AL 35805} }

\begin{abstract}
Nonthermal radiation observed from astrophysical systems containing
relativistic jets and shocks, e.g., gamma-ray bursts (GRBs), active
galactic nuclei (AGNs), and microquasars commonly exhibit
power-law emission spectra. Recent PIC simulations of relativistic
electron-ion (or electron-positron) jets injected into a stationary
medium show that particle acceleration occurs within the downstream
jet. In collisionless, relativistic shocks, particle (electron,
positron, and ion) acceleration is due to plasma waves and their
associated instabilities (e.g., the Weibel (filamentation) instability) 
created in the shock region. The simulations show that
the Weibel instability is responsible for generating and amplifying
highly non-uniform, small-scale magnetic fields. These fields
contribute to the electron's transverse deflection behind the jet
head. The resulting ``jitter'' radiation from deflected electrons has 
different properties compared to synchrotron radiation, which assumes a uniform 
magnetic field. Jitter radiation may be important for understanding the
complex time evolution and/or spectra in gamma-ray bursts,
relativistic jets in general, and supernova remnants.
\end{abstract}

\maketitle

\vspace{-0.4cm} 
\section{Introduction}

\vspace{-0.3cm} 
Shocks are believed to be responsible for prompt emission from gamma-ray 
bursts (GRBs) and their afterglows, for variable emission from blazars, 
and for particle acceleration processes in jets from active galactic
nuclei (AGN) and supernova remnants (SNRs). The predominant contribution
to the observed emission spectra is often assumed to be synchrotron- and 
inverse Compton radiation from these accelerated particles \cite{p05a,zangr07}.
It is assumed that turbulent magnetic fields in the shock region 
lead to Fermi acceleration, producing higher energy particles \cite{fermi49,blaneich97}.
To make progress in understanding 
emission from these object classes, it is essential to place modeling efforts
on a firm physical basis. This requires studies of the microphysics of the 
shock process in a self-consistent manner \cite{p05b,wax06}. 

\vspace{-0.5cm} 
\subsection{New Numerical Method for Calculating Emission}

\vspace{-0.3cm} 

The retarded electric field from a charged particle moving with
instantaneous velocity $\boldsymbol{\beta}$ under acceleration
$\boldsymbol{\dot{\beta}}$ is obtained \cite{jackson99,nishi08a,nishi08b,nishi08c}. 
After some calculation and simplifying assumptions  the total energy $W$ radiated per unit
solid angle per unit frequency can be expressed as
\vspace*{-0.3cm}
\begin{eqnarray}
&  &\frac{d^{2}W}{d\Omega d\omega}  =  \\ \nonumber
 & & \frac{\mu_{0} c
q^{2}}{16\pi^{3}} \left|\int^{\infty}_{\infty}\frac{\bf{n}\times
[(\bf{n}-\boldsymbol{\beta})\times \boldsymbol{\dot{\beta}}]}{(1-\boldsymbol{\beta}\cdot
\bf{n})^{2}} e^{i\omega(t^{'} -\bf{n} \cdot \bf{r}_{0}({\rm t}^{'})/{\rm c})}
dt^{'}\right|^{2}
\end{eqnarray}

\vspace{-0.2cm} 
Here, $\bf{n} \equiv \bf{R}(\rm{t}^{'})/ |\bf{R}(\rm{t}^{'})|$ is a
unit vector that points from the particle's retarded position towards
the observer. The first term on the right hand side, containing the
velocity field, is the Coulomb field from a charge moving without
influence from external forces in eq. 2.4 \cite{hedeT05}. The second term is a correction term
that arises when the charge is subject to acceleration. Since the
velocity-dependent field falls off in distance as $R^{-2}$, while the
acceleration-dependent field scales as $R^{-1}$, the latter becomes
dominant when observing the charge at large distances ($R \gg 1$).
The choice of unit vector $\bf{n}$ along the direction of propagation of
the jet (hereafter taken to be the $x$-axis) corresponds to head-on emission. 
For any other choice of $\bf{n}$ (e.g., $\theta = 1/\gamma$), off-axis emission 
is seen by the observer. The observer's viewing angle is set by the choice of 
$\bf{n}$ ($n_{\rm x}^{2}+n_{\rm y}^{2}+n_{\rm z}^{2} = 1$). 

%



\vspace{-0.6cm} 
 \begin{table}
\begin{tabular}{lrrrrrrrr}
\hline
  & \tablehead{1}{r}{b}{$B_{\rm x}$}
  & \tablehead{1}{r}{b}{$V_{\rm j1,2}$}
  & \tablehead{1}{r}{b}{$V_{\perp,1}$}   
  & \tablehead{1}{r}{b}{$V_{\perp,2}$}
  & \tablehead{1}{r}{b}{$\gamma_{\max}$}
  & \tablehead{1}{r}{b}{$\theta_{\Gamma}$}
  & \tablehead{1}{r}{b}{Remarks}   
  
  \\
\hline

\hline
P   &  3.70 ($B_{\rm z}$)  & 0.0c          &  0.998c  & 0.9997c & 40.08   &4.491   & gyrating\\
 A &  3.70                          & 0.99c       &   0.1c     &  0.12c    & 13.48   &13.35   & jet\\ 
 B &  3.70                        & 0.9924c   &   0.1c      &  0.12c    & 36.70    &4.905  & jet\\
 C &  3.70                          & 0.99c       &   0.01c   &  0.012c   & 7.114   &25.30    & jet\\
 D &  0.370                        & 0.99c       &   0.01c    &  0.012c  & 7.114   &25.30   & jet\\
 E &  0.370                        & 0.99c       &   0.1c     &  0.12c    & 13.48   &13.35  & $\Delta t = 0.005$\\
 F &  0.370                       & 0.99c        &   0.1c      &  0.12c    & 13.48  &13.35    & $\Delta t = 0.025$\\

\hline
\end{tabular}
\caption{Seven cases of radiation}
\label{tab:b}
\end{table}

\begin{figure}[!th]
\centering
\vspace*{-1.2cm}
\includegraphics[width=0.41\linewidth]{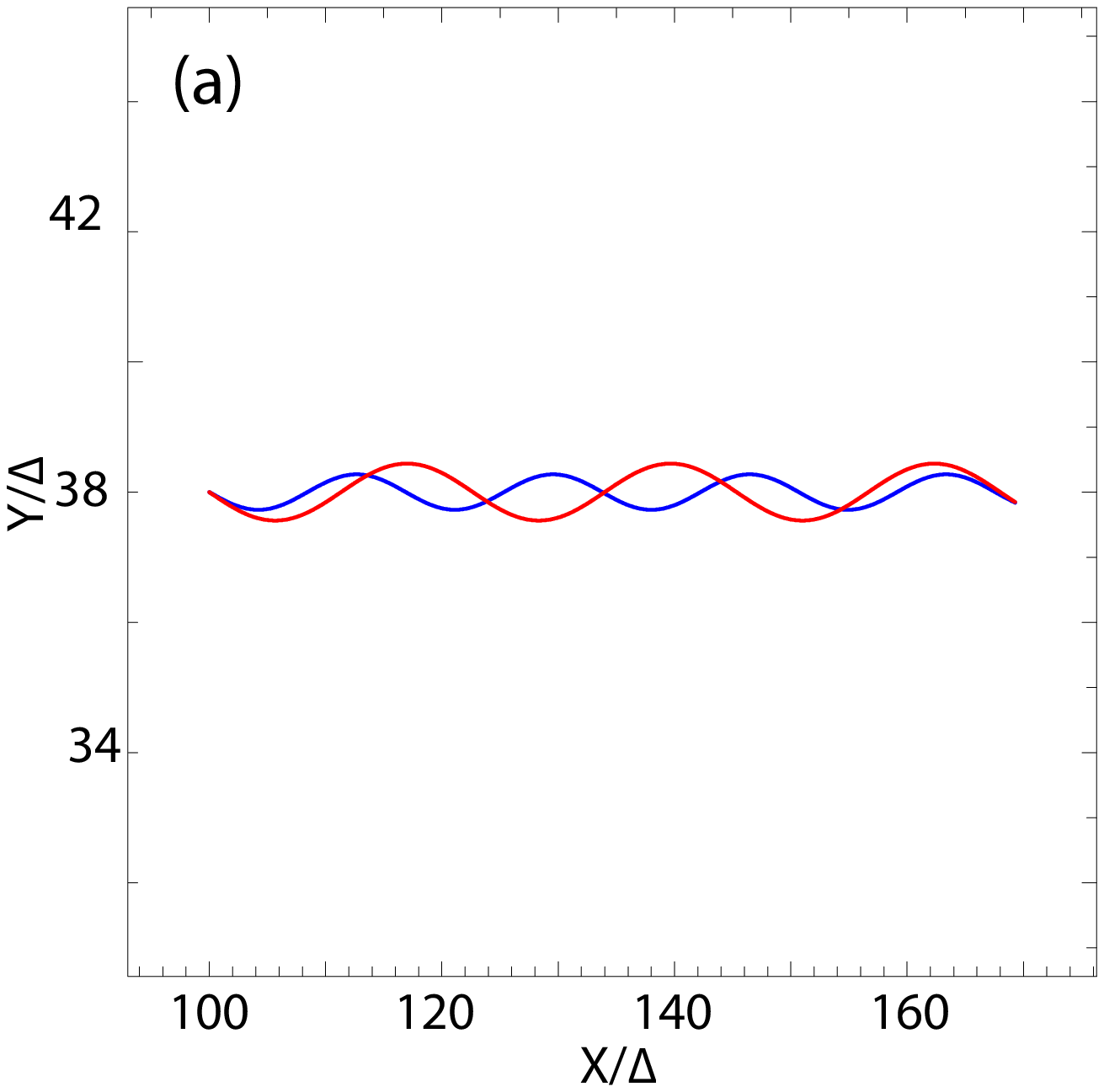}
\includegraphics[width=0.83\linewidth]{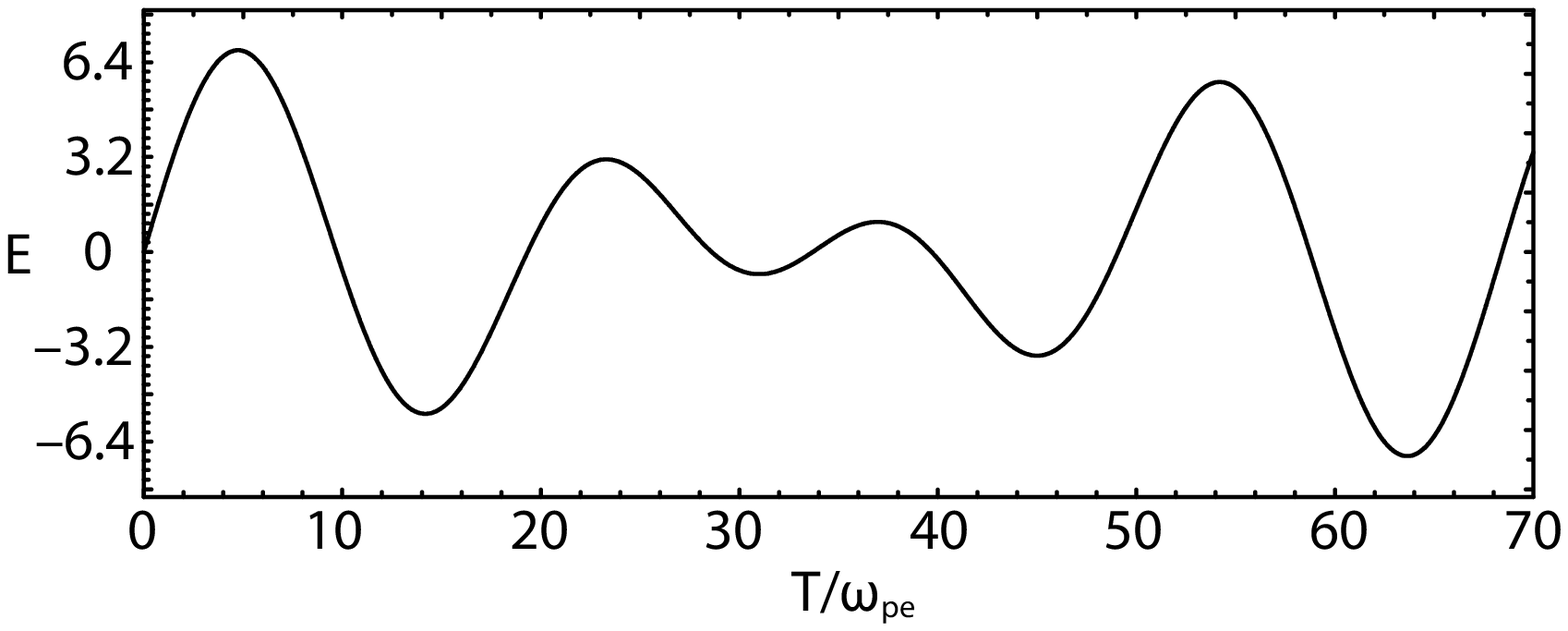}
\includegraphics[width=0.41\linewidth]{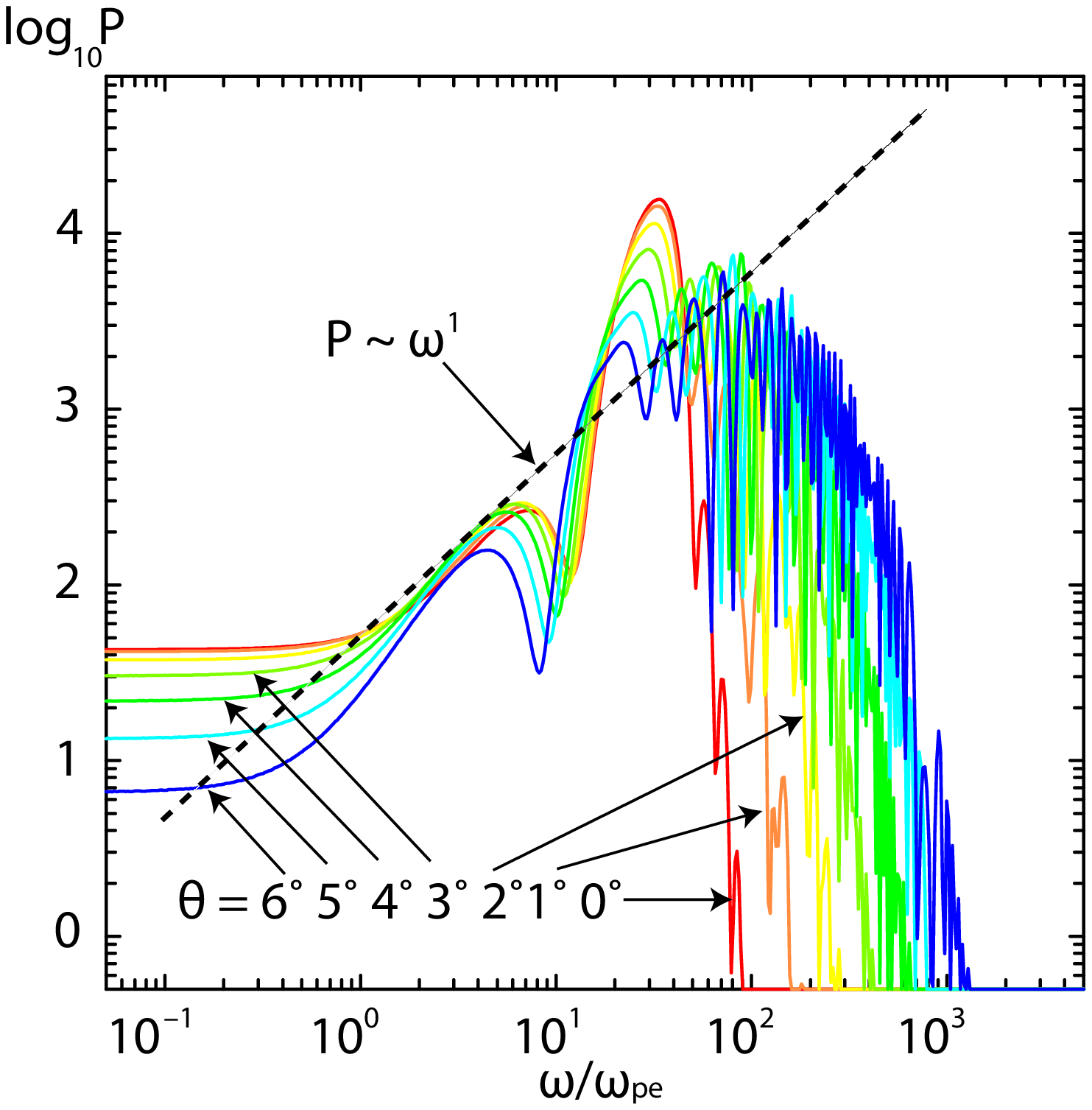}
\end{figure}
\begin{figure}[!th]
\centering
\vspace*{-1.6cm}
\includegraphics[width=0.41\linewidth]{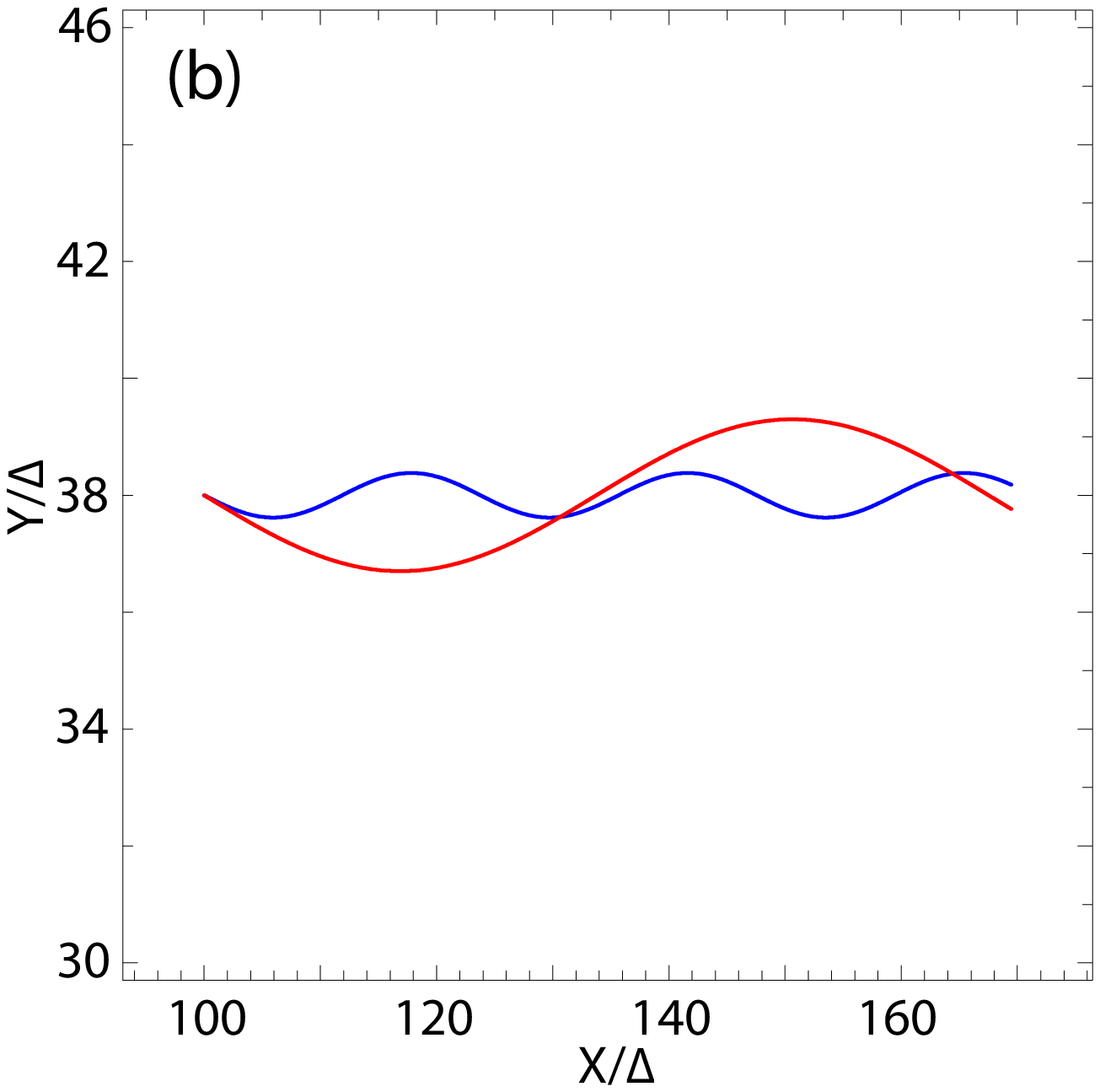}
\includegraphics[width=0.83\linewidth]{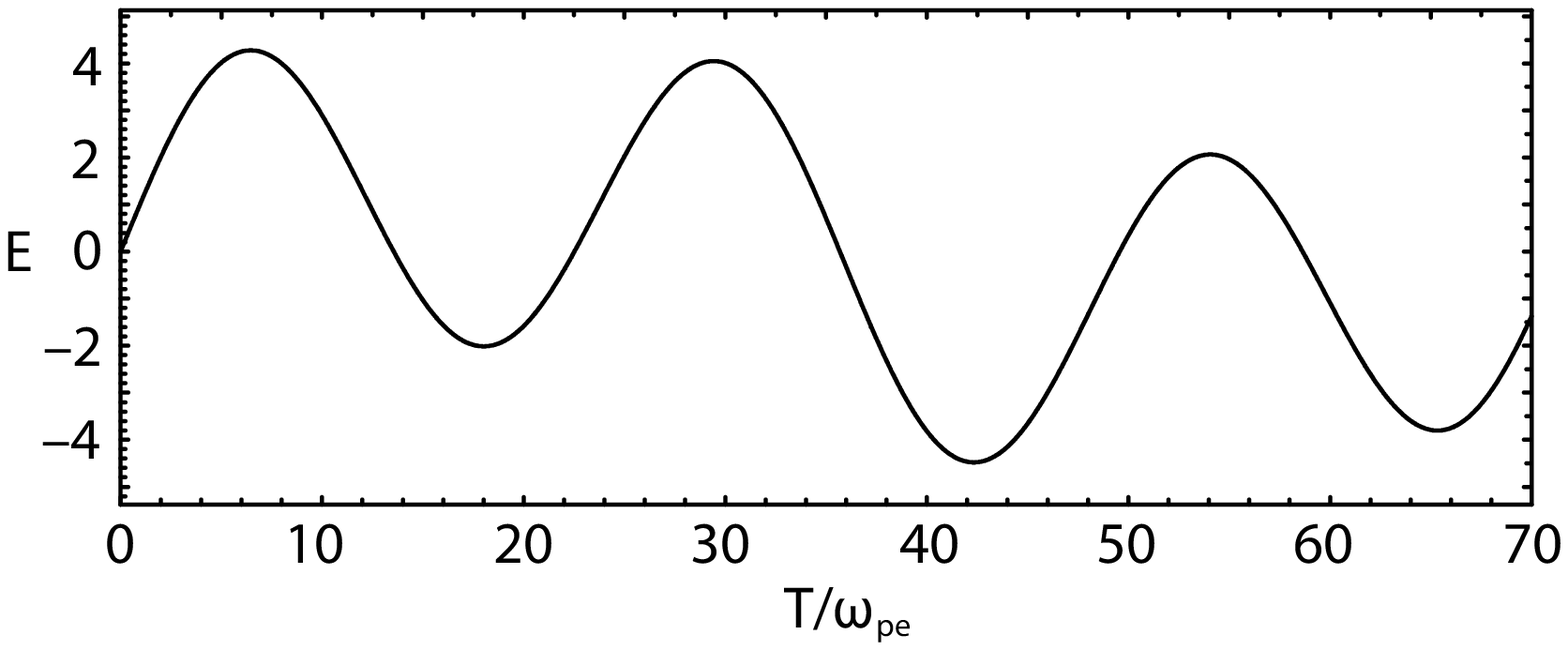}
\includegraphics[width=0.41\linewidth]{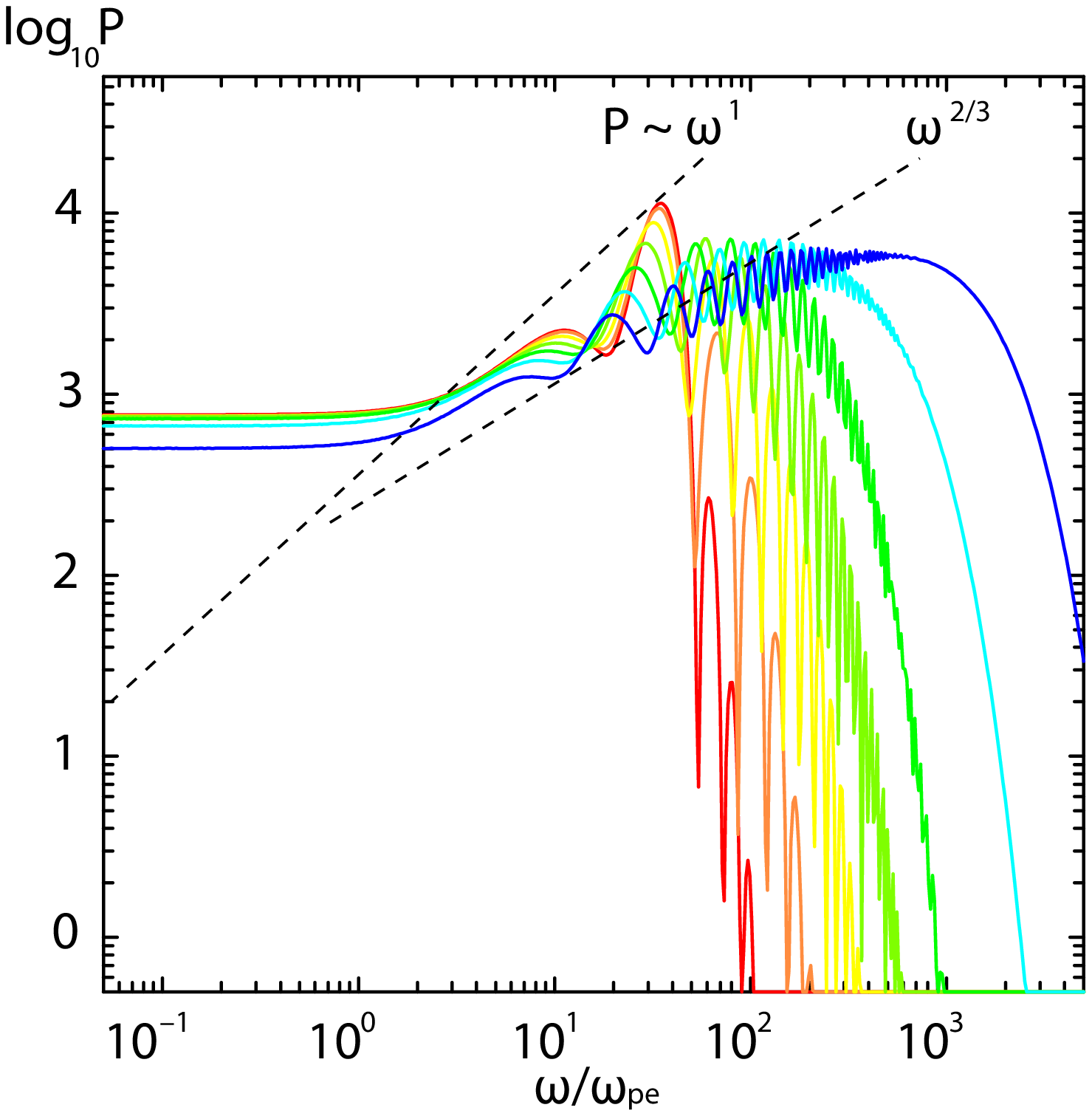}
\end{figure}
\begin{figure}[!th]
\centering
\vspace*{-1.6cm}
\includegraphics[width=0.41\linewidth]{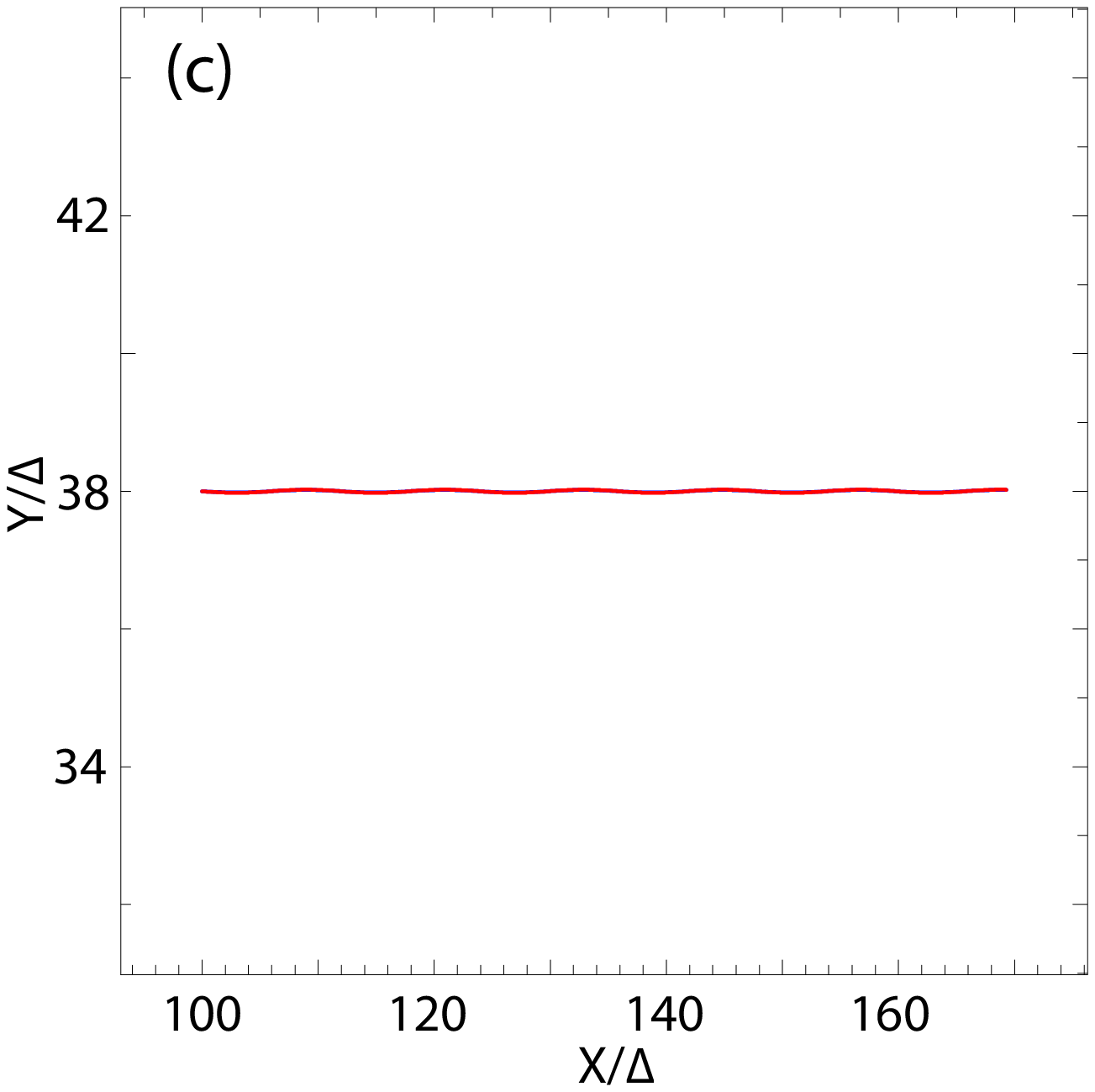}
\includegraphics[width=0.83\linewidth]{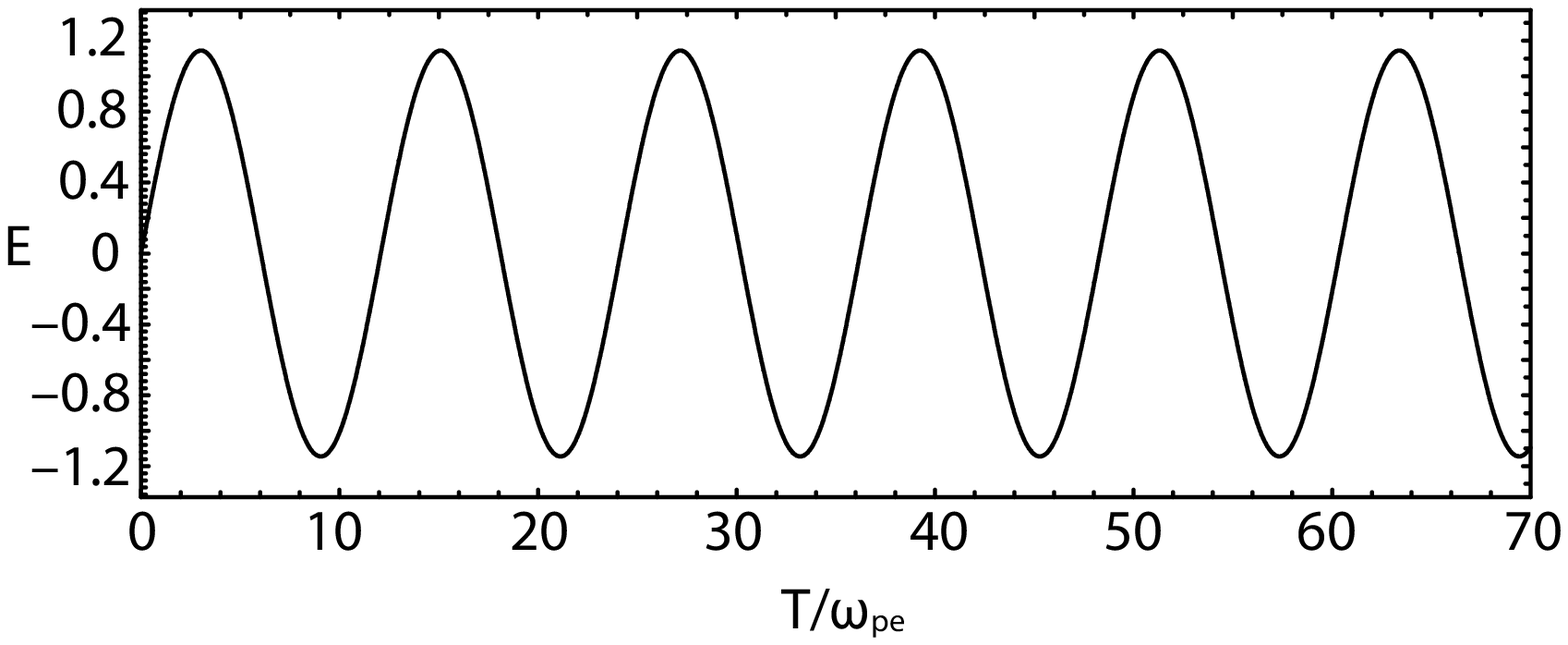}
\includegraphics[width=0.41\linewidth]{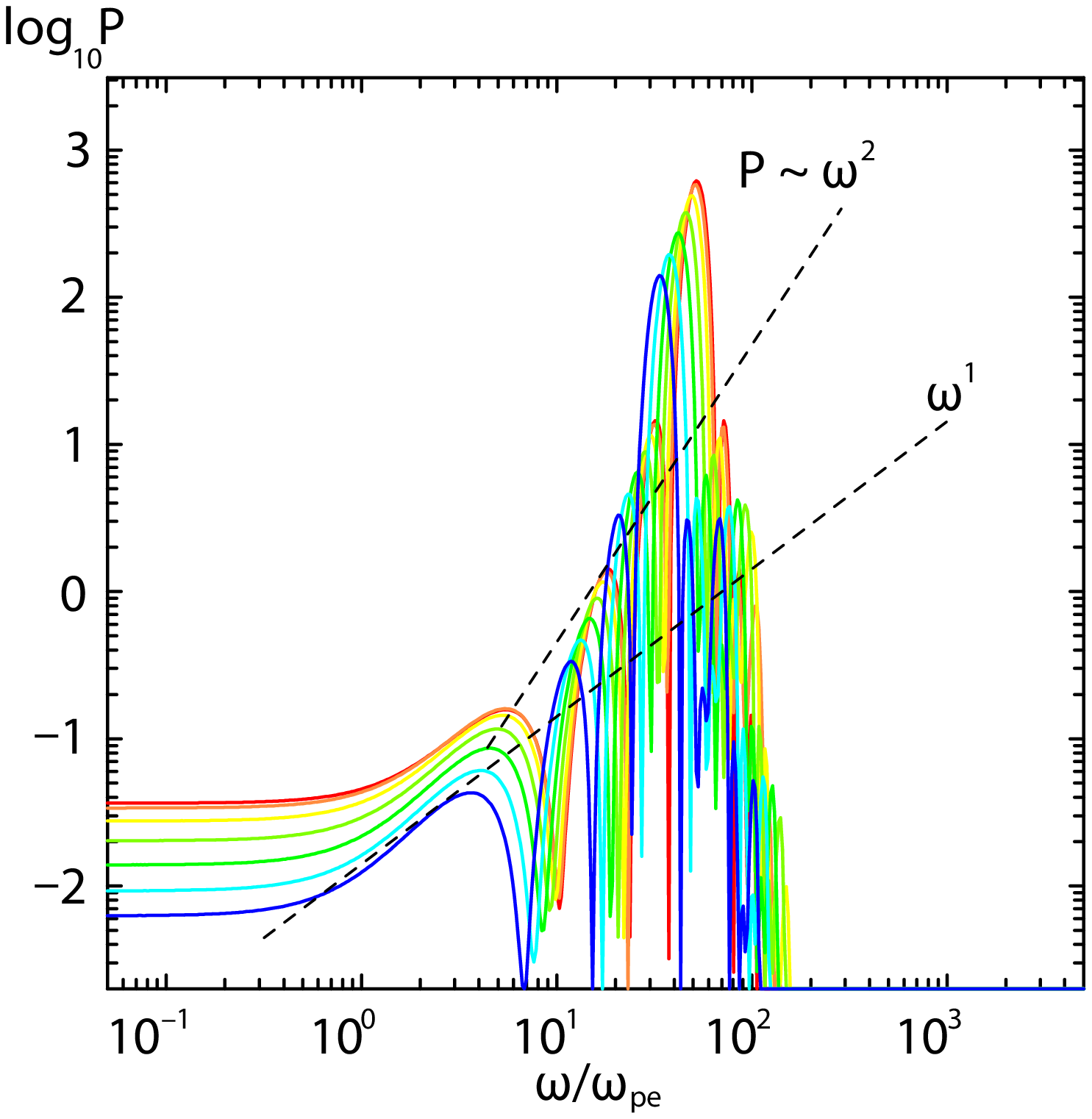}
\end{figure}
\begin{figure}[!th]
\centering
\vspace*{-1.6cm}
\includegraphics[width=0.41\linewidth]{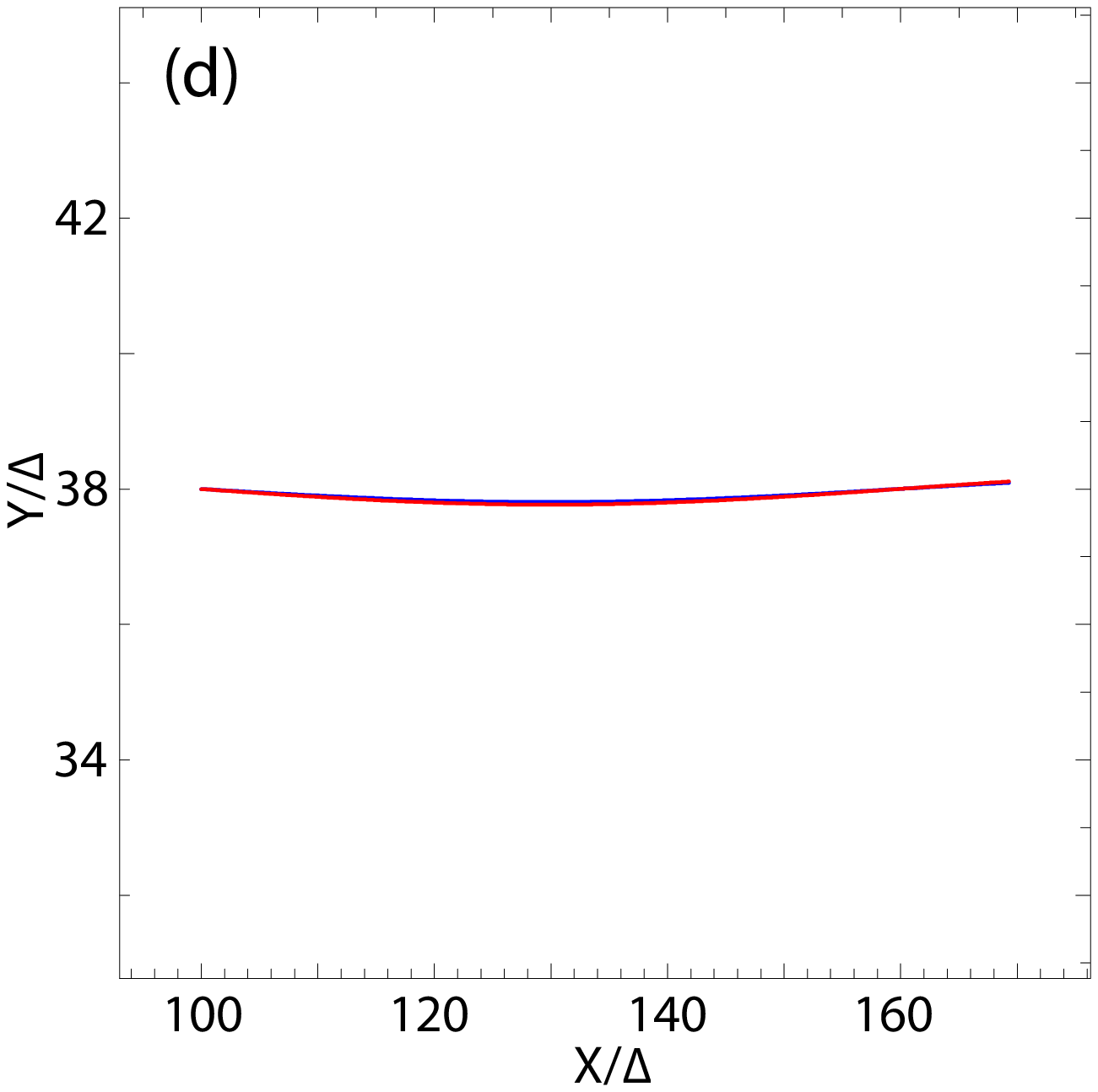}
\includegraphics[width=0.83\linewidth]{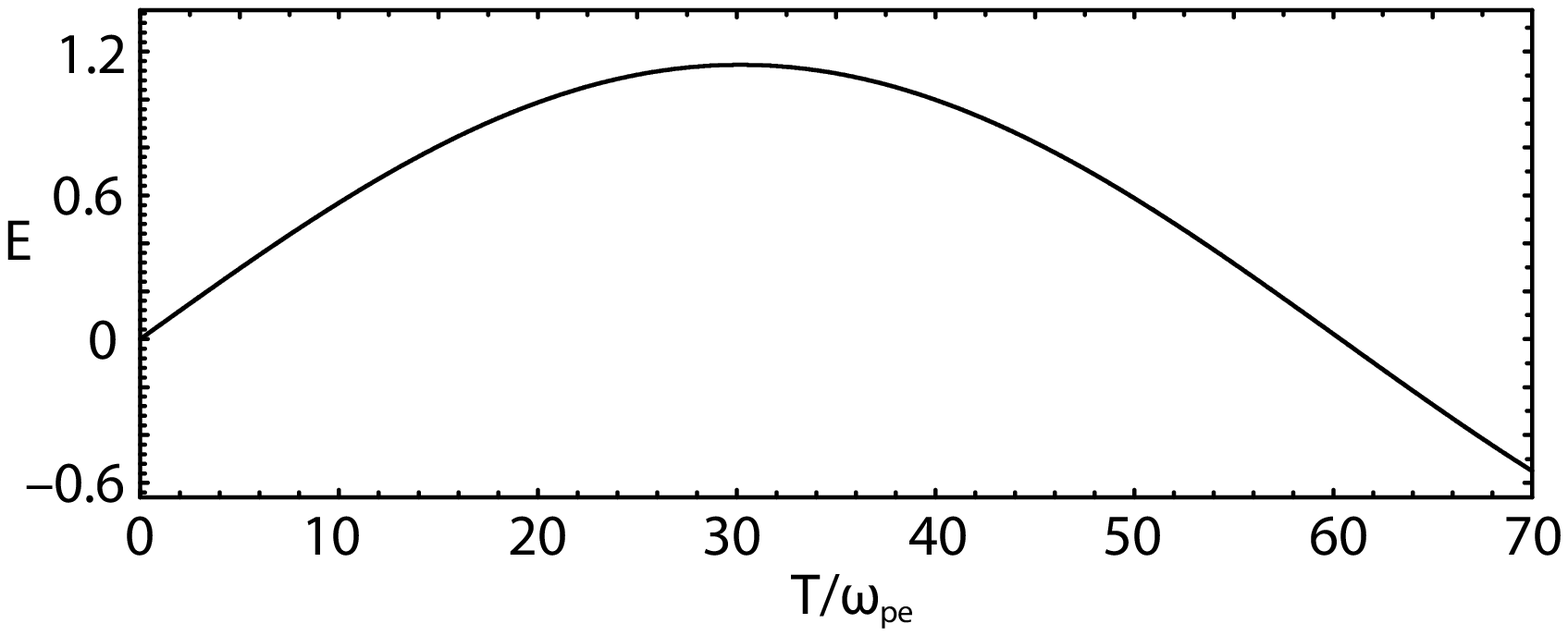}
\includegraphics[width=0.41\linewidth]{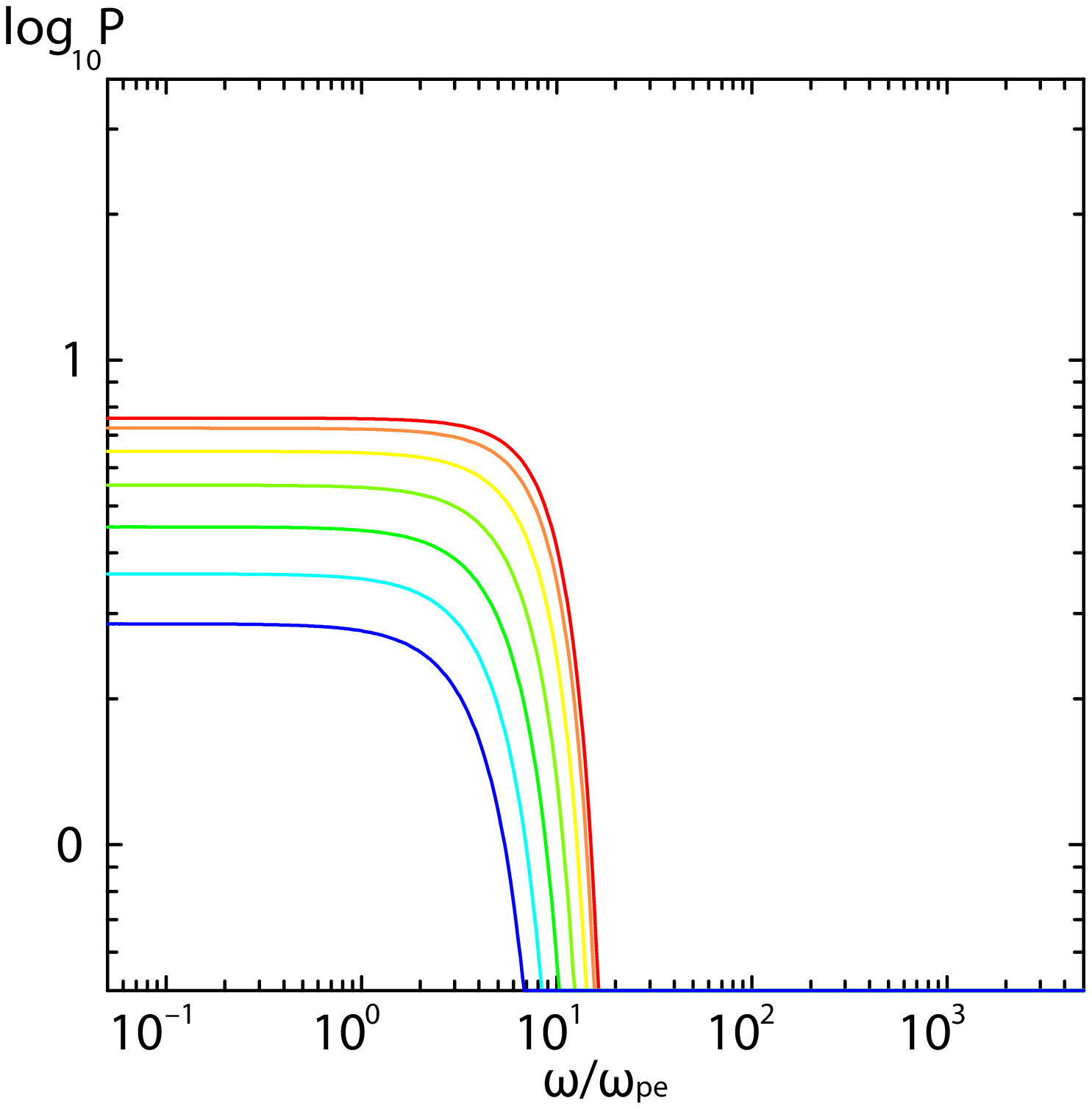}
\end{figure}
\begin{figure}[!th]
\centering
\vspace*{-1.6cm}
\includegraphics[width=0.41\linewidth]{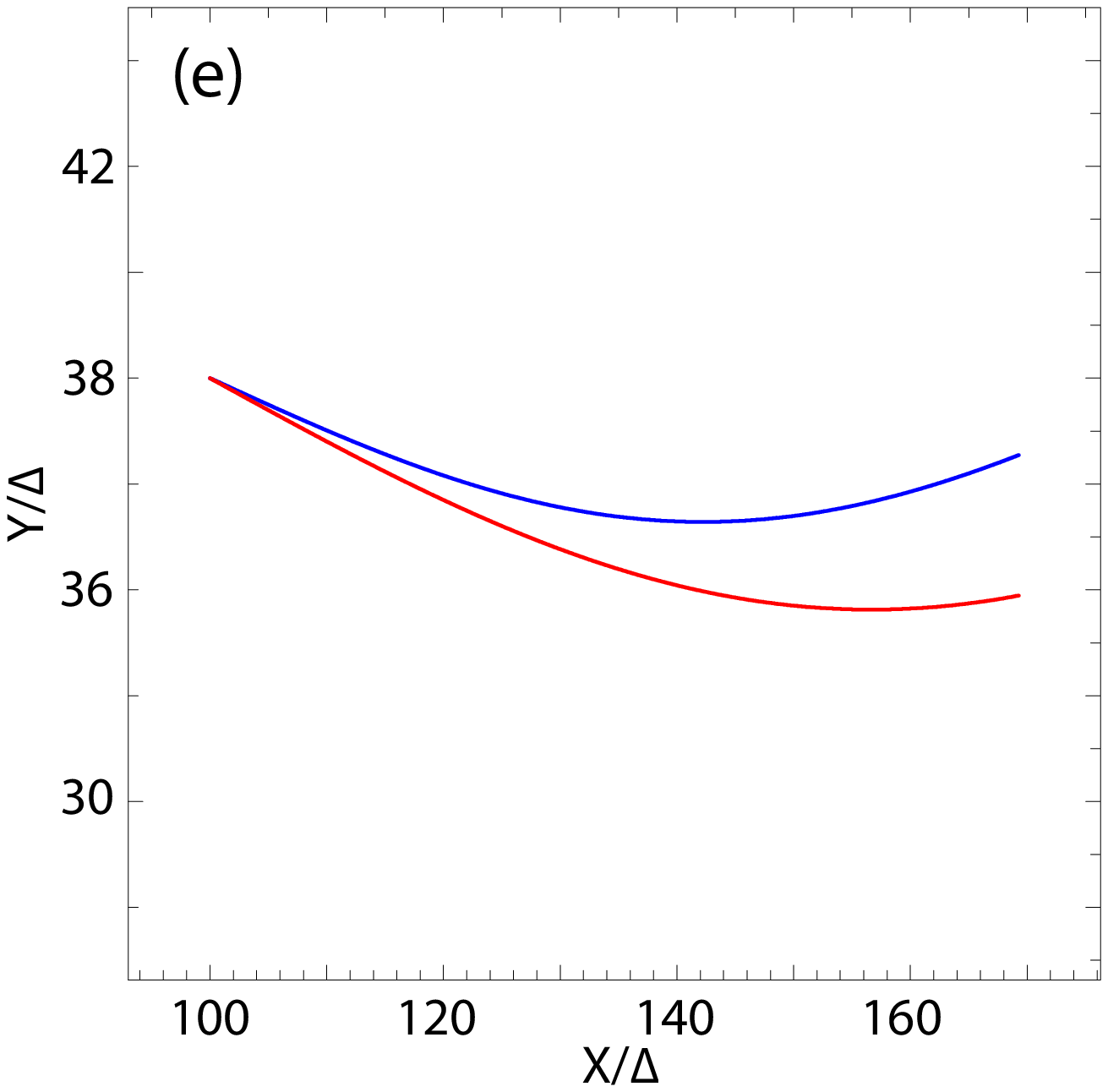}
\includegraphics[width=0.83\linewidth]{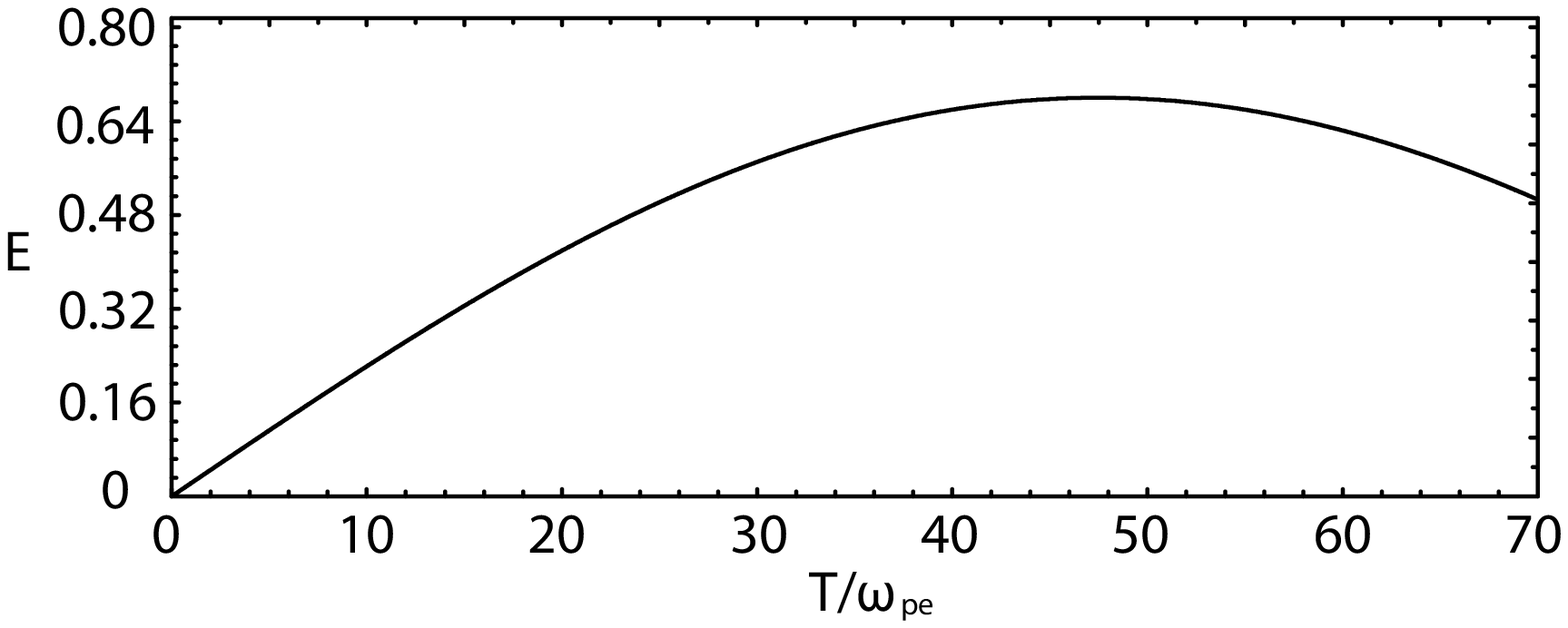}
\includegraphics[width=0.41\linewidth]{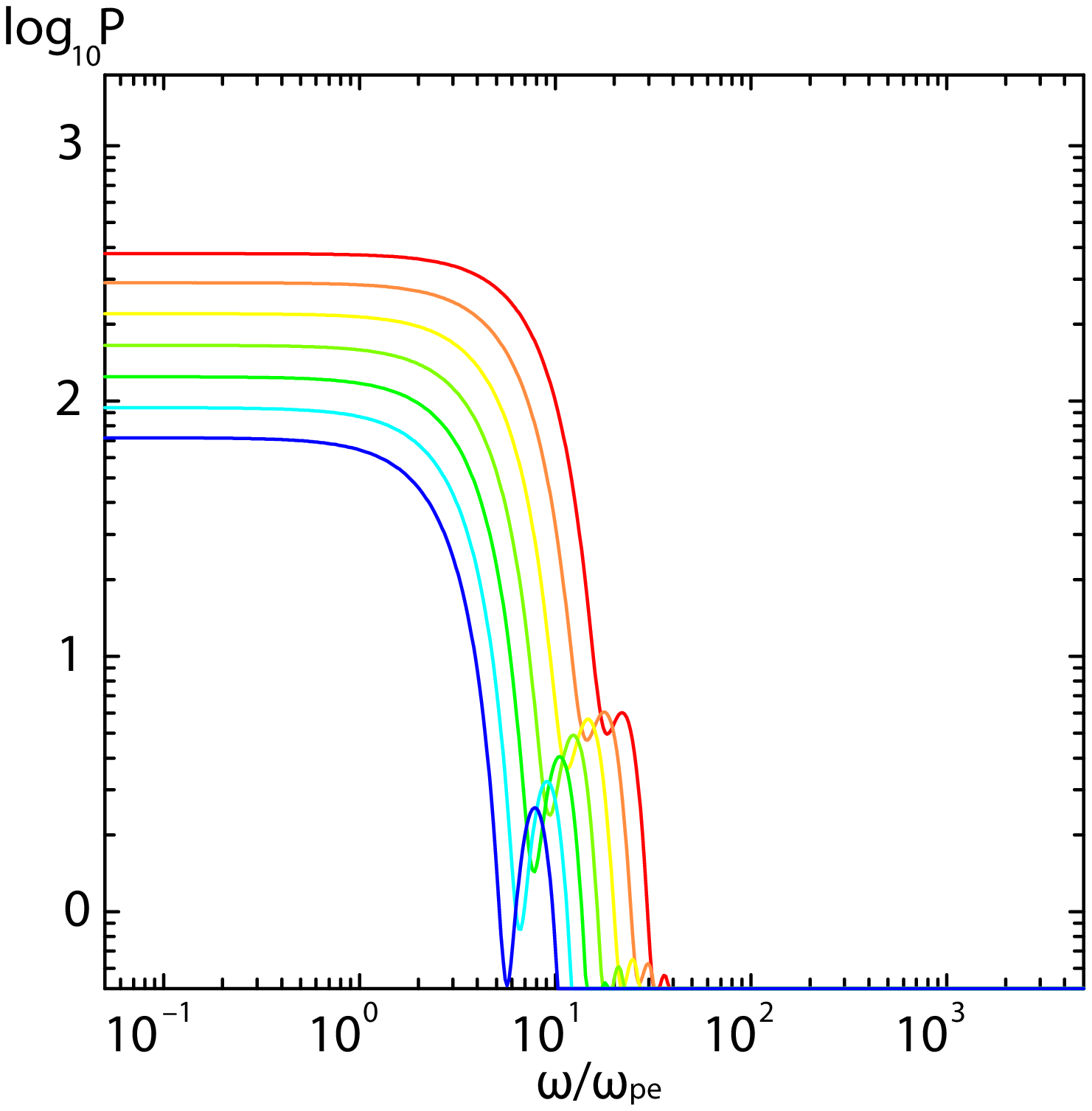}
\end{figure}
\begin{figure}[!th]
\centering
\vspace*{-1.6cm}
\includegraphics[width=0.41\linewidth]{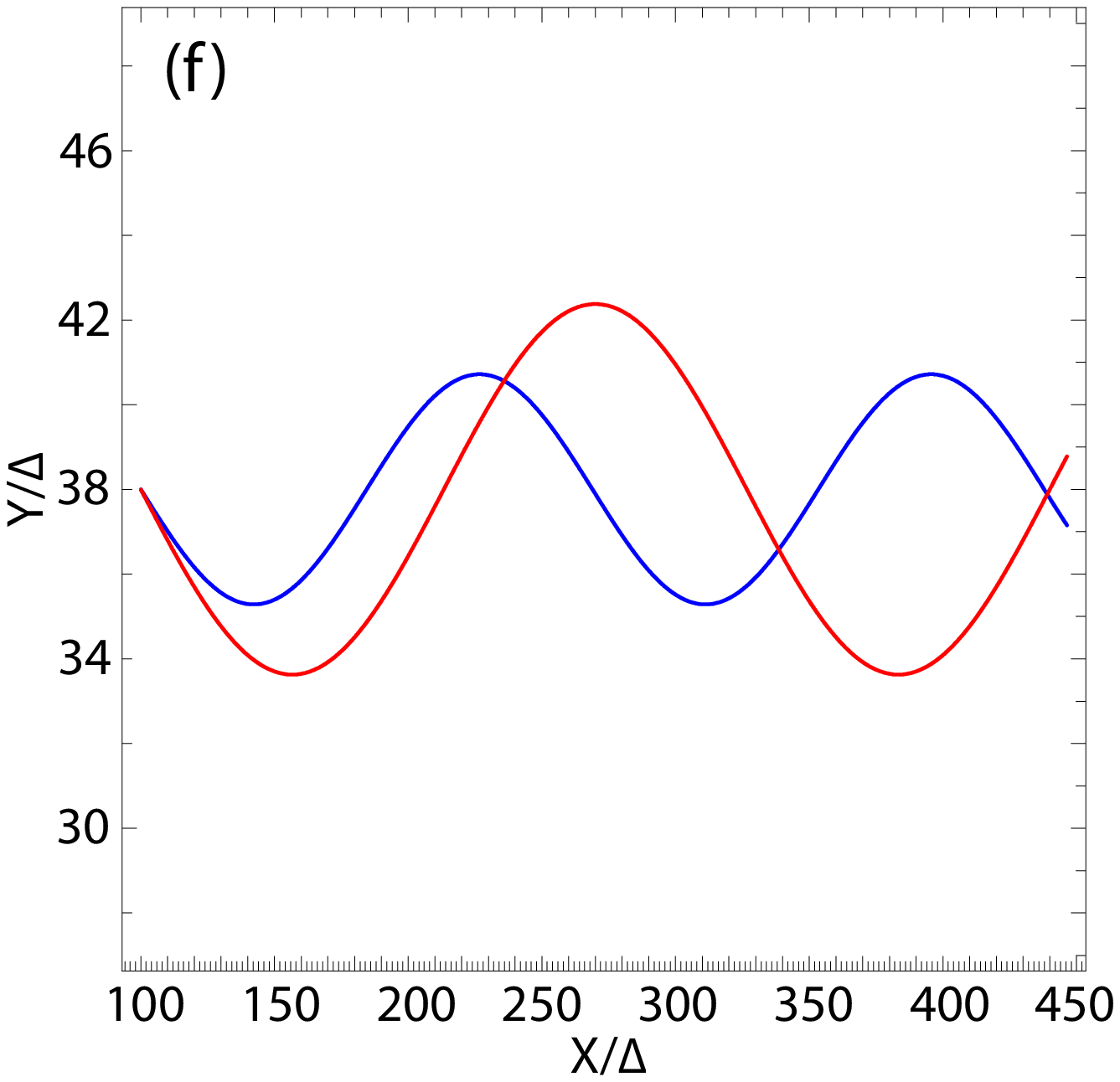}
\includegraphics[width=0.83\linewidth]{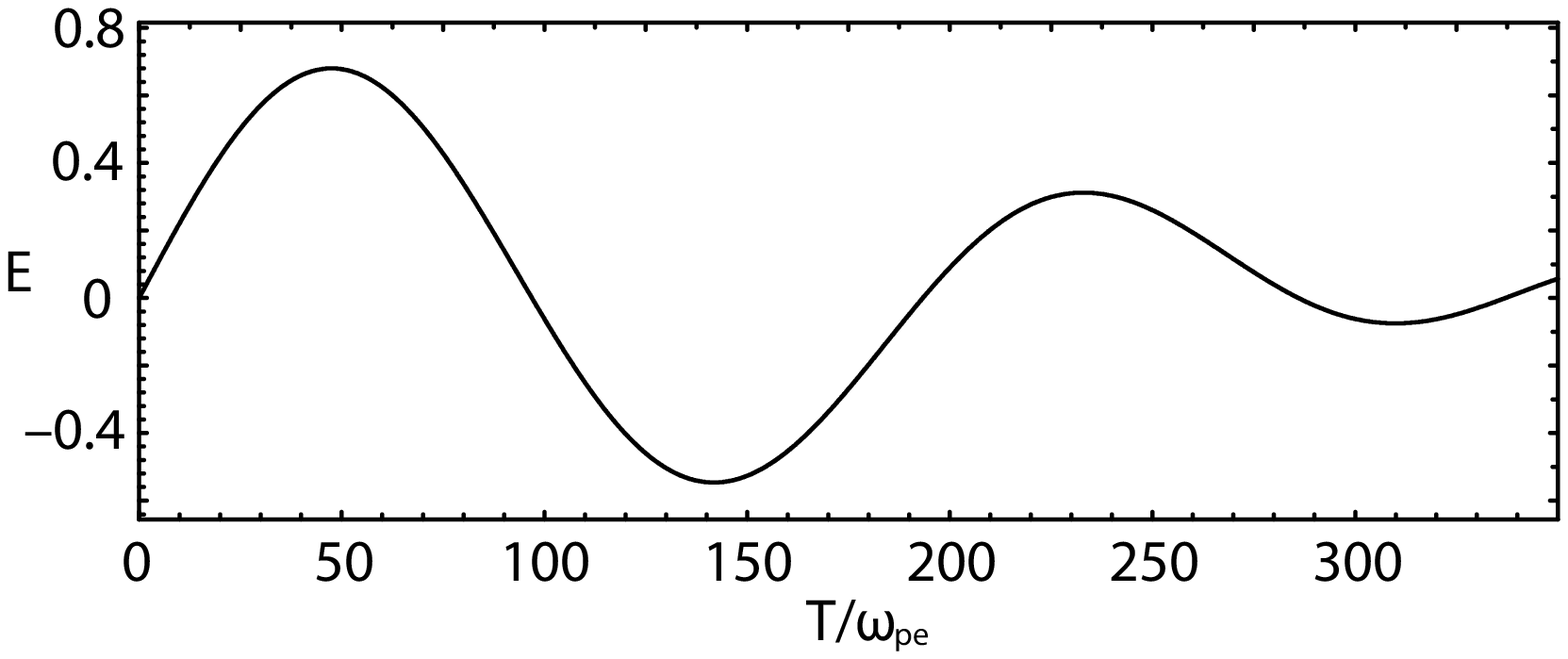}
\includegraphics[width=0.41\linewidth]{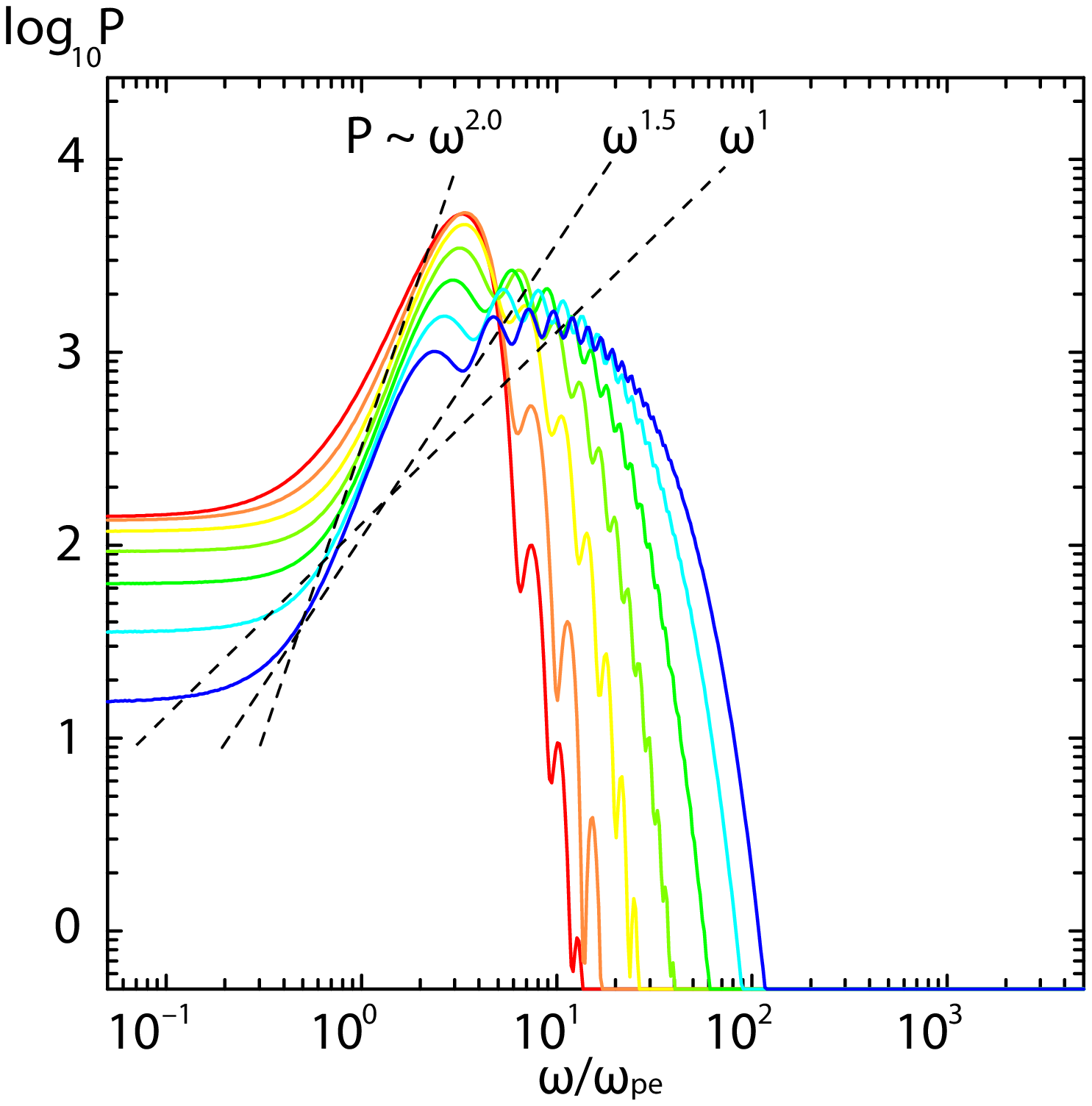}

\caption{Summary of six cases with jet velocity (Cases A - F) (for Case P, see Nishikawa et al. \cite{nishi08c})}
\end{figure}

\vspace*{-0.9cm}
\section{Radiation from two electrons}

\vspace*{-0.3cm}
In the previous section we discussed how to obtain the retarded
electric field from relativistically moving particles (electrons)
observed at large distance. Using eq.\ 1 
we calculated the time
evolution of the retarded electric field and the spectrum from a
gyrating electron in a uniform magnetic field to verify the technique
used in this calculation. We have calculated 
the radiation from two electrons gyrating in the $x - y$ plane in the uniform
magnetic field $B_{\rm z}$ with Lorentz factors ($\gamma = 15.8, 40.8$) (Case P in Table 1)
\cite{nishi08a,nishi08b,nishi08c}.
We have very good agreement between the spectrum obtained from the simulation
and the theoretical synchrotron spectrum expectation  from 
eq. 7.10 \cite{hede05}.

In order to calculate more realistic radiation from relativistic jets we included a parallel magnetic field 
($B_{\rm x}$). Relativistic jets are propagating along the $x$ direction. Table 1 shows six cases including
the previous case P (first row) \cite{nishi08c}.   
The jet velocity is 0.99c (except Case B). Two different magnetic field strengths are used. Two electrons 
are injected with two different perpendicular velocities (Cases A - F). The maximum Lorenz factors, 
$\gamma_{\max} =\{(1 - (V_{\rm j2}^{2} +V_{\perp 2}^{2})/c^{2}\}^{-1/2} $ are calculated for  
larger perpendicular velocity.  The critical angles for the off-axis radiation is calculated
with  $\theta_{\Gamma} = \Gamma^{-1}$.

Figure 1 shows the summary of the six cases. Trajectories of the two electrons are shown in the left column 
(red: larger perpendicular velocity, blue: smaller perpendicular velocity).
The two electrons propagate from left to right with gyration in the $y - z$ plane (not shown).  The gyroradius 
is about  $0.44\Delta$ ($\Delta = 1$: the simulation grid length) for the electron with a larger perpendicular velocity (Case A).  The radiation 
electric field from the two electrons is shown in the middle column.
The spectra were calculated  at the point $(x, y, z) = (64,000,000.0, 43.0, 43.0)$ shown in he right column. 
The seven curves show the spectrum at the viewing 
angles 0$^{\circ}$ (red), 1$^{\circ}$ (orange), 2$^{\circ}$ (yellow), 3$^{\circ}$ (moss green),
4$^{\circ}$ (green), 5$^{\circ}$ (light blue), and 6$^{\circ}$ (blue) ($n_{\rm y}\ne 0$). The higher frequencies become stronger with the increasing viewing angle. For Case A the power spectrum is scaled as $P \sim \omega^{1}$
as proposed for jitter radiation \cite{medv06}. For all Cases the spectra are much steeper than 
the slope $1/3$ for the synchrotron radiation.

The second row in Fig. 1 shows Case B with a larger jet velocity $V_{\rm j1,2} = 0.9924c$ with the 
other parameters  kept the same as Case A. The spectra with larger 
viewing  angles are similar to those of Case A. The spectrum slope is smaller than that in Case A.
 However, due to the large jet velocity the higher frequencies
at larger viewing angles (0$^{\circ}$, 1$^{\circ}$, 2$^{\circ}$) become stronger. On the other hand, with the 
smaller perpendicular velocities (Cases C and D), the gyroradius becomes very small. The spectra become weaker than
those in Case A. As shown in the third and fourth rows in Fig. 1, the viewing angle dependence becomes very small. It should be noted that the slope of spectra is very steep for Case C. Spectral leakage is found. 

Cases D - F  have a weaker magnetic fields ($B_{\rm x} = 0.370$)  than Cases A - C. Case
D  has a small perpendicular velocity. The trajectories are almost
straight. The spectra look very similar to that for Bremsstrahlung \cite{hede05}. The spectra
become flat at lower frequencies. The peak spectral power is the weakest of all
the cases. 
With larger perpendicular velocities the spectra become stronger than those with the smaller perpendicular velocities. Case F shows the case with a larger time step (5 times) with the same parameters as Case E. The spectrum slope is very steep.
The spectra in this case show two differences with those in Case E. First there exist positive slopes in the lower
frequency. Second, due to the gyro-motion the spectra split due to the viewing angles. In particular the spectrum
with  larger viewing angle becomes stronger at high frequencies.  

As shown in Table 1, the critical angles for the off-axis radiation $\theta_{\Gamma} = \Gamma^{-1}$
are different. In this study we have obtained the off-axis radiation for the angles
0$^{\circ}$, 1$^{\circ}$, 2$^{\circ}$, 3$^{\circ}$,
4$^{\circ}$, 5$^{\circ}$, and 6$^{\circ}$ ($n_{\rm y}\ne 0$). For cases D, E, and F 
the variation among different viewing angles is small since the angles are much smaller than 
the critical angle (25.3$^{\circ}$ and 13.35$^{\circ}$). However, for case B ($\theta_{\Gamma} = 5^{\circ}$)
the radiation shows larger differences for different viewing angles due to the small critical angle. For Case F (a longer time 
($340/\omega_{\rm pe}$) the spectra at high frequencies  become stronger with larger viewing angles. 

These results validate the technique used in our code. It should be noted 
that the method based on the integration of the retarded electric fields calculated
by tracing many electrons described in the previous section can provide a
proper spectrum in turbulent electromagnetic fields. On the other
hand, if the formula for the frequency spectrum of radiation emitted
by a relativistic charged particle in instantaneous circular motion
is used \cite{jackson99,rybic79}, the complex particle
accelerations and trajectories are not properly accounted for and the
jitter radiation spectrum is not properly obtained.

\vspace{-0.4cm} 
\section{Discussion}

\vspace{-0.3cm} 
The procedure used to calculate jitter radiation using the technique
described in the previous section has been implemented in our code.

In order to obtain the spectrum of synchrotron (jitter) emission \cite{medv00,medv06,flei06}, we
consider an ensemble of electrons randomly selected in the region
where the filamentation (Weibel) instability \cite{weib59} has fully developed, and
electrons are accelerated in the generated magnetic fields. We
calculate emission from about 20,000 electrons during the sampling
time, $t_{\rm s} = t_{\rm 2} - t_{\rm 1}$ with Nyquist frequency
$\omega_{\rm N} = 1/2\Delta t$ where $\Delta t$ is the simulation time
step and the frequency resolution $\Delta \omega = 1/t_{\rm
s}$. However, since the emission coordinate frame for each particle is
different, we accumulate radiation at fixed angles in simulation
system coordinates after transforming from the individual particle
emission coordinate frame. This provides an intensity spectrum as a
function of angle relative to the simulation frame $x$-axis (this can 
be any angle by changing the unit vector ${\bf n}$ in eq.\ (1)). A
hypothetical observer in the ambient medium (viewing the external GRB
shock) views emission along the system $x$-axis. This computation is
carried out in the reference frame of the ambient medium in the
numerical simulation. For an observer located outside the direction
of bulk motion of the ambient medium, e.g., internal jet shocks in an
ambient medium moving with respect to the observer, an additional
Lorentz transformation would be needed along the line of sight to the
observer. Spectra obtained from simulations can be rescaled to physical time scales. 

Emission obtained by the method described above is self-consistent,
and automatically accounts for magnetic field structures on the
small scales responsible for jitter emission. By performing such
calculations for simulations with different parameters, we can then
investigate and compare the quite contrasted regimes of jitter- and
synchrotron-type emission \cite{medv00,medv06,flei06} for prompt and afterglow 
emission. 
The feasibility of this approach has been demonstrated and implemented
\cite{hedeN05,hede05}. 
Thus, we will be able to
address the issue of low frequency GRB spectral index violation of the
synchrotron line of death \cite{medv06}. 

Simulations incorporating jitter radiation are in progress using an MPI code \cite{niem08} which speeds up considerably from the previous OpenMP code \cite{ram07}. New results of jitter radiation will be presented separately.

\vspace{-0.4cm} 
\begin{theacknowledgments}

\vspace{-0.3cm} 
We have benefited from many useful discussions with 
A. J. van der Horst. This work is supported by AST-0506719, 
AST-0506666, NASA-NNG05GK73G and NNX07AJ88G. JN was supported by MNiSW research 
projects 1 P03D 003 29 and N N203 393034,
 and The Foundation for Polish Science through the HOMING program, which is
 supported through the EEA Financial Mechanism.Simulations were performed at the  
Columbia facility at the NASA
Advanced Supercomputing (NAS).  
Part of this work was done while K.-I. N. was visiting the
Niels Bohr Institute. He thanks the director of the institution for 
generous hospitality.

\end{theacknowledgments}

\end{document}